\def\aa{{A\&A}}

\def\aj{{AJ}}
\def\annrev{{ARA\&A}}
\def\apj{{ApJ}}
\def\apjs{{ApJS}}

\def\mnras{{MNRAS}}

\def\pasp{{PASP}}

\documentclass[cup5b]{caps}
\usepackage{graphicx}
\usepackage{amssymb}
\usepackage{ociwsymp1}
\HeadText{T. M. Heckman}

\begin{document}

\pagenumbering{arabic}

\author[]{TIMOTHY M. HECKMAN\\Center for Astrophysical Sciences, Department
of Physics \& Astronomy,\\The Johns Hopkins University} 

\chapter{Star Formation in Active Galaxies: \\ A Spectroscopic Perspective}

\begin{abstract}

I review the relationship between star formation and black hole building, 
based on spectroscopic observations of the stellar population in active 
galactic nuclei (AGNs) and their host galaxies.  My emphasis is on large, 
well-defined local samples of AGNs, whose optical continua are dominated by 
starlight.  I summarize the spectroscopic tools used to characterize
their stellar content.  Surveys of the nuclei of the nearest galaxies show 
that the stellar population in most low-power AGNs is predominantly old.  In 
contrast, young ($< 1$ Gyr) stars are detected in about half of the powerful 
type 2 Seyfert nuclei. I summarize Sloan Digital Sky Survey spectroscopy of 
the host galaxies of 22,000 AGNs.  The AGN phenomenon is commonplace only in
galaxies with high mass, high velocity dispersion, and high stellar surface
mass density. Most normal galaxies with these properties have old stellar 
populations. However, the hosts of powerful AGNs (Seyfert 2s) have young 
stellar populations.  The fraction of AGN hosts that have recently undergone 
a major starburst also increases with AGN luminosity.  A powerful AGN 
presumably requires a massive black hole (host with a substantial bulge) plus 
an abundant fuel supply (a star-forming ISM). This combination is rare today, 
but would have been far more common at early epochs.
  
\end{abstract}

\section{Introduction}

\subsection{Motivation}

There is no doubt that that there is a profound physical connection
between the creation of supermassive black holes and the formation
of bulges and elliptical galaxies. The remarkably tight correlation
between the stellar velocity dispersion and black hole mass in these
systems (Ferrarese \& Merritt 2000; Gebhardt .et al. 2000) provides
powerful evidence for this connection. The similarities and differences 
between the strong cosmic evolution of active galactic nuclei (AGNs)
and star formation (e.g., Steidel et al. 1999; Fan et al. 2001) then lead 
naturally to the speculation that AGN evolution traces the build up of the
spheroidal component of galaxies (e.g., Haehnelt \& Kauffman 2000; Kauffmann 
\& Haehnelt 2000).

To put the connection between black hole and galaxy building in context, it is 
interesting to note that the ``universal'' ratio of $\sim10^{-3}$ between 
the black hole and stellar mass in spheroids implies that a moderately powerful
AGN with a bolometric luminosity of $10^{11} \,L_{\odot}$ fueled by accretion 
with radiative efficiency of 10\% would require an associated average 
star formation rate of 65 $M_{\odot}$ yr$^{-1}$ if the spheroid and black hole 
were to be built on the same time scale.  The young stars would outshine 
the AGN by nearly an order of magnitude in this case!

Several interrelated questions immediately arise, and have indeed formed
the basis for most of the discussion at this conference.  What are the 
relevant astrophysical processes that connect the formation of bulges and 
black holes? What conditions foster the fueling of a powerful AGN? Given our 
current understanding of galaxy evolution, can we account for the cosmic 
evolution of the AGN population? Can we see a scaled-down version of 
bulge building surrounding powerful AGNs in the local Universe, or, instead, 
are powerful AGNs today just the temporary rejuvenation of a pre-existing 
black hole with little associated star formation? Can we make any sense of the 
complex gas/star/black hole ecosystem with all its messy gastrophysics?

\subsection{My Perspective}

My goal is to review what is presently known about the possible connection 
between star formation and the AGN phenomenon, both in active galaxies and in 
their nuclei. This is sometimes called the ``starburst-AGN connection'' (e.g., 
Terlevich 1989; Heckman 1991), a vast, sprawling topic that cannot be 
adequately reviewed in a single short paper. Excellent overviews may be found 
in the volumes edited by Filippenko (1992) and Aretxaga, Kunth, \& Mujica 
(2001a). Other recent reviews that complement the present paper include those 
by Veilleux (2001) and Cid Fernandes, Schmitt, \& Storchi-Bergmann (2001a).

This paper will be focused in scope and methodology.  I will just review 
observations, and only discuss spectroscopic measures of the stellar 
populations (ignoring indirect evidence for star formation like molecular gas, 
and far-IR, UV, and radio continuum emission). I will strongly emphasize 
systematic investigations of large well-defined samples. These restrictions 
will allow me to make rather robust statements about the empirical basis for 
a connection between AGNs and star formation. However, these restrictions
effectively limit me to the local Universe.  While the black hole/bulge 
connection was largely established at high redshift, I hope to show that we 
can gain key insights from the fossil record and from local analogs.

\subsection{Spectral Diagnostics of Young Stars}

For young massive stars the strongest spectral features with the greatest 
diagnostic power lie in the vacuum UV regime between the Lyman break and 
$\sim$2000 \AA\ (e.g., Leitherer et al. 1999; de Mello, Leitherer, \& Heckman 
2000). These include the strong stellar wind lines of the O~VI, N~V, Si~IV, 
and C~IV resonance transitions and a host of weaker stellar photospheric 
lines.  Most of the photospheric lines arise from highly excited states and 
their stellar origin is unambiguous. While resonance absorption lines may have 
an interstellar origin, the characteristic widths of the stellar wind profiles 
make them robust indicators of the presence of massive stars. Unfortunately,
observations in this spectral regime are difficult. Only a handful of local 
type 2 Seyferts and LINERs have nuclear UV fluxes that are high enough to 
enable a spectroscopic investigation. While this small sample may not be 
representative, the available data firmly establish the presence of a dominant 
population of young stars (Heckman et al. 1997; Gonz\'alez Delgado et al. 
1998; Maoz et al. 1998; Colina \& Arribas 1999).

While the optical spectral window is far more accessible, the available 
diagnostic features of massive stars are weaker and less easy to interpret.  
Old stars are cool and have many strong spectral features in the optical due 
to molecules and low-ionization metallic species.  Hot young stars have 
relatively featureless optical spectra\footnote{The most direct optical 
signatures of young massive stars are the photospheric lines of He~I 
and the broad emission features due to Wolf-Rayet stars. The former
reach peak strength in early B stars and hence in stellar populations
with ages of tens of Myr (Gonz\'alez Delgado, Leitherer, \& Heckman 1999).
The latter trace the most massive stars and reach peak strengths
for ages of several Myr (e.g., Leitherer et al. 1999). Both the He~I 
and Wolf-Rayet features are weak and contaminated by nebular emission lines.
They have been detected in only a handful of AGNs (Heckman et al. 1997;
Gonz\'alez Delgado et al. 1998; Storchi-Bergmann, Cid Fernandes, \& Schmitt
1998).}.  Thus, the spectroscopic impact of the presence of young stars is 
mostly an indirect one: as they contribute an increasing fraction of the light 
(as the luminosity-weighted mean stellar age decreases) most of the strongest 
spectral features in the optical weaken. This effect is easy to measure. 
Unfortunately, the effect of adding ``featureless'' nonstellar continuum
from an AGN and young starlight will be similar in this regard.

The strongest optical absorption lines from young stars are the Balmer lines. 
These reach peak strength in early A-type stars, and so they are most 
sensitive to a stellar population with an age of $\sim$100 Myr to 1 Gyr (e.g., 
Gonz\'alez Delgado et al. 1998).  Thus, the Balmer lines do not uniquely trace 
the youngest stellar population\footnote{In view of this, and in view of the 
importance of the Balmer lines, throughout this review I will use the term 
``young'' to mean stellar populations with ages less than a Gyr.}.  On the 
plus side, they can be used to characterize past bursts of star formation 
(e.g., Dressler \& Gunn 1983; Kauffmann et al. 2003a).

The situation is summarized in Figure 1.1, which plots the strength of the 
Balmer H$\delta$ absorption line (the H$\delta_A$ index) vs. the 4000 \AA\ 
break strength [the $D_n(4000)$ index] for a set of 32,000 model galaxies with 
different star formation histories (see Kauffmann et al. 2003a).  For 
``well-behaved'' (continuous) histories of past star formation there is a 
tight inverse relation between these two parameters. The addition of a 
starburst moves the galaxy below and to the left of the main locus at early 
times ($\leq$ 100 Myr) times and above it at intermediate times (100 Myr to 
1 Gyr).

\begin{figure}
\centering
\vspace{5mm}
\includegraphics[height=110mm,width=110mm]{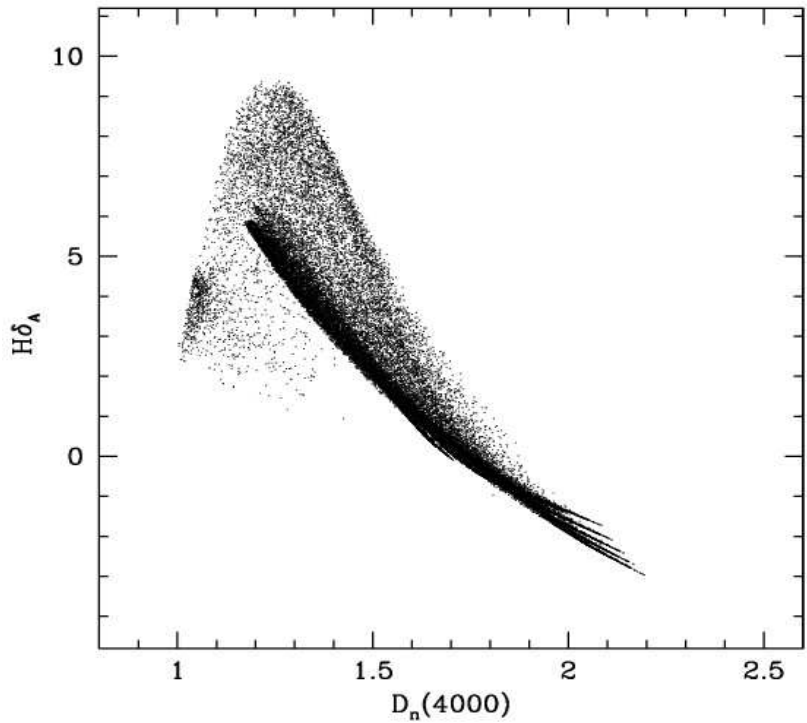}
\caption{Indices measuring the 4000 \AA\ break and H$\delta$ absorption line
are plotted for a grid of 32,000 model galaxies with different star formation
histories. The dark diagonal line is an age sequence of galaxies with
continuous star formation histories (young at upper left and old at lower
right). The other points represent models in which a strong burst of
star formation has occurred within the last $\sim 1$ Gyr. (See Kauffmann
et al. 2003a.)}
\end{figure}

Note that the effect of adding featureless AGN continuum will carry an old 
stellar population located at roughly (2, $-$2) in Figure 1.1 to (1, 0). This 
``mixing line'' lies well under the loci of the stellar populations [the 
latter having much stronger H$\delta_A$ at a given $D_n(4000)$].  This 
underscores the importance of the Balmer absorption lines, and of the need 
to properly account for the contaminating effects of nebular Balmer emission 
lines. The relative nebular contamination is minimized for the high-order 
Balmer lines in the blue and near-UV.

So far, I have equated young stars with hot stars. This is certainly true for 
main-sequence stars. However, a population of red supergiants will contribute 
significantly to the near-IR light in young stellar populations\footnote{For 
an instantaneous burst of star formation, red supergiants do not appear until 
$\sim$5 to 6 Myr have elapsed. Thus, they are absent only in very young bursts 
of very short duration.}.  The spectral features produced by red supergiants 
are qualitatively similar to those produced by red giants that dominate the 
near-IR light in an old stellar population. A robust method to determine 
whether old giants or young supergiants dominate is to measure the $M/L$ ratio 
in the near-IR using the stellar velocity dispersion. So far this technique 
has been applied to only a small sample, but the results are tantalizing 
(Oliva et al. 1999; Schinnerer et al. 2003). 

In an old stellar population, cool stars provide most of the light in the 
optical and near-IR and so the associated metallic and molecular spectral 
features are strong in both bands. In contrast, the optical (near-IR) 
continuum in a young stellar population will be dominated by hot main-sequence 
stars (cool supergiants). The metallic/molecular lines are therefore weak in 
the optical and strong in the near-IR. This combination of properties provided
some of the first direct evidence for a young stellar population in AGNs
(Terlevich, D\'\i az, \& Terlevich 1990; Nelson \& Whittle 1999).

\subsection{Emission-Line Diagnostics and AGN Classification}

As noted above, I will be emphasizing optical spectroscopy in this review
of the stellar populations in AGNs.  Historically, this has also been the most 
widely used technique to detect and classify AGNs themselves. 

AGNs can be broadly classified depending upon whether the central black hole 
and its associated continuum and broad emission-line region is viewed directly 
(a ``type 1'' AGN) or is obscured by a surrounding dusty medium (a ``type 2'' 
AGN). Since this obscuring medium does not fully cover 4$\pi$ steradians as 
seen from the central AGN.  some of the AGN radiation escapes the central 
region and photoionizes surrounding circumnuclear ($\sim10^2$ to $10^3$ 
pc-scale) gas, leading to strong, relatively narrow permitted and forbidden 
emission lines from the ``narrow-line region'' (NLR). In type 1 AGNs, the 
optical continuum is dominated by nonstellar emission, making it difficult to 
study the stellar population. In what follows, I will therefore ignore this 
type of AGN. In type 2 AGNs the observed optical continuum is predominantly 
starlight, with some contribution by light from the obscured AGN scattered 
into our line of sight and from nebular continuum associated with the NLR
(e.g., Tran 1995; Wills et al. 2002).

In the simplest version of the ``unified'' model for AGNs (Antonucci 1993) the 
type 1 and type 2 AGNs are drawn from the same parent population and differ 
only in our viewing angle.  In this case the stellar content of the two types 
will be the same.  On the other hand, if the solid angle covered by the dusty 
obscuring medium varies substantially, then type 1 (2) AGNs will be 
preferentially drawn from those objects with smaller (larger) covering 
fractions for this medium.  In such a case systematic differences might exist 
in stellar content between type 1 and type 2 AGNs (especially if the dusty 
obscurer is related to star formation).  There have been recurring suggestions 
to this effect (e.g., Maiolino et al. 1995; Malkan, Gorjian, \& Tam 1998; 
Oliva et al. 1999).  It is important to keep this possible bias in mind.

Narrow emission lines can also be produced via photoionization by hot young 
stars. The most unambiguous discrimination between excitation of the NLR by 
young stars vs. an AGN is provided by lines due to species with ionization 
potentials greater than 54 eV (above the He~II edge), since a young stellar 
population produces a negligible supply of such high-energy photons. 
Unfortunately, most such high-ionization lines are weak in the optical spectra 
of AGNs. In practice, classification is therefore usually based on the flux 
ratios of the strongest lines (Heckman 1980a,b;  Baldwin, Phillips, \& 
Terlevich 1981; Veilleux \& Osterbrock 1987).

\begin{figure}
\centering
\includegraphics[height=120mm,width=120mm]{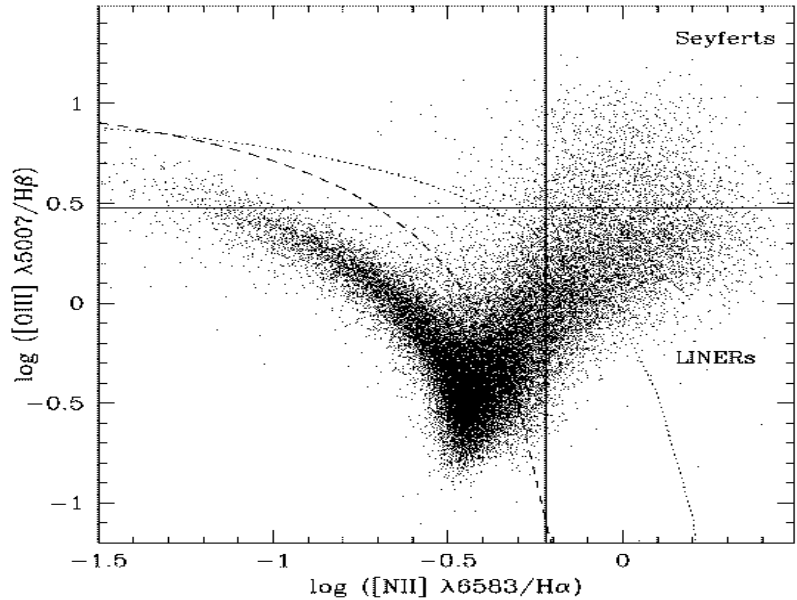}
\caption{Diagnostic flux-ratio diagram for a sample of nearly 56,000 
emission-line galaxies from the SDSS.  According to Kewley et al. (2001) 
galaxies dominated by an AGN will lie above and to the right of the dotted 
line.  Galaxies lying below and to the left of the dashed line are dominated
by star-forming regions.  Galaxies lying between the dashed and dotted curves 
are transition objects (AGN and star formation are both important). The 
locations of classic AGN-dominated Seyfert nuclei and LINERs are indicated.}
\end{figure}

An example of such a classification diagram is shown in Figure 1.2, based on 
nearly 56,000 emission-line galaxies from the Sloan Digital Sky Survey (SDSS) 
sample discussed below. In this plot of the [O~III] $\lambda$5007/H$\beta$ vs. 
the [N~II] $\lambda$6583/H$\alpha$ NLR line ratios, two distinct sequences are 
present. Star-forming galaxies define a narrow locus in which the metallicity 
increases from the upper left to the lower center. The plume that ascends from
the high-metallicity end of the starformer sequence toward higher values of 
both [O~III]/H$\beta$ and [N~II]/H$\alpha$ is the AGN population. In simple 
physical terms, the high-energy continuum of an AGN results in much greater 
photoelectric heating per ionization, which raises the temperature in the 
ionized gas, and therefore strengthens the collisionally excited forbidden 
lines that cool the gas. 

It is important to realize that a contribution to the emission-line spectrum 
by regions of star formation is almost inevitable in many of the AGNs in 
Figure 1.2: at the median redshift of the SDSS main galaxy sample 
($z \approx 0.1$)
the 3$^{\prime\prime}$ diameter SDSS fiber corresponds to a projected 
size of nearly 6 kpc. Thus, a galaxy's position along the AGN plume will be 
determined by the relative contribution of the AGN and star formation to the 
emission-line spectrum.  This is particularly important in the context of this 
review because it means that the signature of young stars is present in the 
nebular emission lines as well as in the stellar absorption lines.

The AGN plume is quite broad at its upper end, where the AGN contribution is 
dominant. Traditionally, these AGN-dominated objects are classified as 
Seyferts if [O~III]/H$\beta >$ 3 and as low-ionization nuclear emission-line 
regions (LINERs) if [O~III]/H$\beta <$ 3 (e.g., Veilleux \& Osterbrock 1987). 
The latter AGNs are more common, but do not attain the high luminosities of 
powerful Seyfert nuclei (e.g., Heckman 1980b; Ho, Filippenko, \& Sargent 2003).

\section{Young Stars in Active Galactic Nuclei}

Spectroscopy of the nearest AGNs affords the opportunity to study the 
starburst-AGN connection on small physical scales ($>$ a few pc).  The 
drawback is that the nearest AGNs have low luminosities, and we might expect 
that the amount of star formation associated with black hole fueling would 
scale in some way with AGN power. To get a complete picture it is thus 
important to examine both the nearest AGNs and more powerful AGNs. I will 
review these two regimes in turn.

The earliest investigation of the stellar population for a moderately large 
sample of the nearest AGNs was by Heckman (1980a,b), who discussed 30 LINERs 
found in a survey of a sample of 90 optically bright galaxies.  The typical 
projected radius of the spectroscopic aperture was $\sim$200 pc.  The spectra 
covered the range from 3500 to 5300 \AA. LINERs were primarily found in 
galaxies of early Hubble type (E through Sb). Based on the strengths of the
stellar metallic lines and the high-order Balmer lines, the nuclear continuum 
was dominated by old stars in about 3/4 of the LINERs, while a contribution of 
younger stars was clearly present in the remainder.  Typical luminosities of 
the [O~III] $\lambda$5007 and H$\alpha$ NLR emission lines were 
$\sim 10^5$ to $10^6 \,L_{\odot}$.

Ho et al. (2003) have recently examined the nuclear (typical radius $\sim 100$ 
pc) stellar population in a complete sample of $\sim$500 bright, nearby 
galaxies (of which 43\% contain an AGN).  In nearly all respects this is a 
major step forward from my old analysis. The sample is considerably larger, 
the quality of the spectra is superior, and the analysis of the emission-line
properties more careful and more sophisticated. The larger sample size and 
improved treatment of the emission lines allow Ho et al. to study 
statistically significant samples of low-luminosity LINERs, type 2 Seyferts, 
and transition nuclei, and to span a larger range in AGN luminosity 
($L_{{\rm H}\alpha} \approx 10^4$ to $10^7 \, L_{\odot}$).  The only 
disadvantage of these spectra is that they do not extend shortward of 4230 
\AA\, and so they miss the H$\delta$ and higher-order Balmer lines that most 
effectively probe young stars. Nevertheless, their results are qualitatively 
consistent with those of Heckman (1980a,b). The AGNs are hosted by early-type 
galaxies (E through Sbc). With a few exceptions, the AGNs have predominantly 
old stellar populations.  The exceptions are primarily transition nuclei whose 
emission-line spectra suggest excitation by a mix of an AGN and massive stars.

My colleagues (Cid Fernandes, Gonz\'alez Delgado, Schmitt, Storchi-Bergmann) 
and I have recently undertaken a program of near-UV spectroscopy of 43 LINER
and transition nuclei taken from the Ho, Filippenko, \& Sargent (1997) survey. 
Our specific goal was to access the high-order stellar Balmer absorption 
lines. We detect these lines in about half of the transition nuclei, but in 
very few LINERs.  The cases in which hot stars are present are primarily Sb, 
Sbc, or Sc galaxies. These results are consistent with the idea that 
transition nuclei are composite AGN/starformers.

Let me now discuss significantly more powerful AGNs (classical type 2 
Seyferts). It was recognized very early-on (e.g., Koski 1978) that the optical 
spectra of powerful type 2 Seyfert nuclei could not be explained purely by an 
old stellar population. An additional ``featureless continuum'' that typically 
produced 10\% to 50\% of the optical continuum was present. Until relatively 
recently, it was tacitly assumed that this component was light from the AGN 
(plausibly light from the hidden type 1 Seyfert nucleus that had been 
reflected into our line of sight by free electrons and/or dust). However, a 
detailed spectropolarimetric investigation by Tran (1995) and related arguments
by Cid Fernandes \& Terlevich (1995) and Heckman et al. (1995) showed that 
only a small fraction of this continuum could be attributed to scattered AGN 
light. While nebular continuum emission must be present (e.g., Tran 1995; 
Wills et al. 2002), Cid Fernandes \& Terlevich (1995) and Heckman et al. 
(1995) argued that young stars were the main contributors.

This has been confirmed by several major optical spectroscopic investigations
of moderately large samples of type 2 Seyfert nuclei (Schmitt, 
Storchi-Bergmann, \& Cid Fernandes 1999; Gonz\'alez Delgado, Heckman, \& 
Leitherer 2001; Cid Fernandes et al. 2001b; Joguet et al. 2001). The latter 
two papers examined the nuclear stellar population in 35 and in 79 type 2 
Seyferts, respectively. The projected aperture sizes range from a few hundred 
pc to about a kpc.  On average, these are considerably more powerful AGNs than 
those in the Heckman (1980b) or Ho et al. (2003) samples, with [O~III] 
luminosities of $10^6$ to $10^9 \,L_{\odot}$.

The principal conclusion is that a young ($< 1$ Gyr) stellar population is 
clearly present in about half of the Seyfert 2 nuclei. Cid Fernandes et al. 
(2001b) find that the fraction of nuclei with young stars is $\sim$60\% when 
$L_{\rm [O~III]} > 10^7 \,L_{\odot}$ but is only $\sim$10\% for the less 
powerful nuclei. They also found that the ``young'' Seyfert 2 nuclei were 
hosted by galaxies with much larger far-IR luminosities ($\sim 10^{10}$ to 
$10^{12} \,L_{\odot}$) than the ``old'' nuclei ($\sim 10^{9}$  to $10^{10} 
\,L_{\odot}$), suggesting that the global star formation rate was 
correspondingly higher (a topic considered in some detail below). 
Interestingly, Joguet et al. (2001) found that the Hubble type distribution 
for the host galaxies was roughly the same for the ``old'' and ``young'' 
nuclei (S0 to Sc). Using morphological classifications based on the 
{\it Hubble Space Telescope}\ imaging survey of (Malkan et al. 1998), 
Storchi-Bergmann et al. (2001) find a reasonably good correspondence between 
the presence of a young nuclear population and a late ``inner Hubble type'' 
(the presence of dust lanes and spiral features in the inner few kpc).

To date, spectroscopic investigations of the nuclear stellar populations in
powerful radio-loud type 2 AGNs (the so-called ``narrow-line radio galaxies'')
have been restricted to relatively small samples (Schmitt, Storchi-Bergmann \& 
Cid Fernandes 1999; Aretxaga et al. 2001b; Wills et al. 2002; Tadhunter et al. 
2002). A young nuclear stellar population has been detected in about 1/3 of 
the cases.

\section{Young Stars in AGN Host Galaxies}

In the previous section I have reviewed spectroscopy of the stellar population
in the nuclear region (the centralmost $\sim 10^2$ to $10^3$ pc).  I now turn 
my attention to more global properties of AGN hosts.

Raimann et al. (2003) have recently used long-slit optical/near-UV 
spectroscopy to measure the radial variation in the stellar population for the 
sample of type 2 Seyfert galaxies whose nuclear properties were investigated 
by Gonz\'alez Delgado et al. (2001).  They characterize the stellar content 
by fitting each spectrum with a set of spectra of stellar clusters spanning a 
range in age and metal abundance.

They find that the Seyfert 2's have younger mean ages than their comparison
sample of normal galaxies at both nuclear and off-nuclear locations.  The 
sample-averaged ``vector'' of the fractional contribution to the continuum at 
4000 \AA\ by stellar populations with ages of 10 Gyr, 1 Gyr, 30 Myr, and 3 Myr 
is (0.36, 0.25, 0.24, 0.15) at the nucleus, (0.36, 0.31, 0.20, 0.13) at 1 kpc, 
and (0.32, 0.30, 0.27, 0.11) at 3 kpc.  Thus, the stellar population shows no 
strong systematic radial gradient in the hosts of powerful Seyfert 2 nuclei 
(at least out to radii of several kpc).

\subsection{The Sloan Digital Sky Survey}

So far, I have reported on spectroscopy of modest-sized samples of AGNs.  The 
ongoing SDSS (York et al. 2000; Stoughton et al. 2002; Blanton et al. 2003;
Pier et al. 2003) provides us with the opportunity to investigate the 
structure and stellar content of the hosts of tens of thousands of AGNs! In 
the remainder of the paper I will give a progress report on an on-going 
program in this vein that is a close collaboration between groups at JHU and 
MPA-Garching (Kauffmann et al., in preparation).  The program was made 
possible by the inspiration and perspiration of the dedicated team that over 
the past 13 years has created and operated the SDSS.

The SDSS is using a dedicated 2.5-meter wide-field telescope at the Apache 
Point Observatory to conduct an imaging and spectroscopic survey of about a 
quarter of the sky. The imaging is conducted in the $u$, $g$, $r$, $i$, and 
$z$ bands (Fukugita et al. 1996; Gunn et al. 1998; Hogg et al. 2001; Smith et 
al. 2002), and spectra are obtained with a pair of multi-fiber spectrographs 
built by Alan Uomoto and his team at JHU.  When the survey is complete, 
spectra will have been obtained for nearly 10$^6$ galaxies and 10$^5$ QSOs 
selected from the imaging data. Details on the spectroscopic target selection
for the ``main'' galaxy sample and QSO sample can be found in Strauss et al. 
(2002) and Richards et al. (2002), respectively. We will be summarizing results
based on spectra of $\sim$123,000 galaxies contained in the the SDSS Data
Release One (DR1). These data are to be made publically available early in 2003.

Since I will primarily be discussing results derived from the spectra,
it is useful to summarize their salient features. Spectra are obtained
through 3$^{\prime\prime}$ diameter fibers. At the median redshift of the main
galaxy sample ($z \approx 0.1$) this corresponds to a projected aperture size
of $\sim$6 kpc which typically contains 20\% to 40\% of
the total galaxy light. Thus, the SDSS spectra are closer to global
than to nuclear spectra. At the median redshift the spectra
cover the rest-frame wavelength range from $\sim$3500 to 8500 \AA\
with a spectral resolution $R \approx 2000$ ($\sigma$$_{instr} \approx 65$
~km~s$^{-1}$). The spectra are spectrophotometrically calibrated through
observations of subdwarf F stars in each 3-degree field.
By design, the spectra are well-suited to the determinations
of the principal properties of the stars and ionized gas in galaxies.

\subsection{Our Methodology}

For the convenience of the reader we give a brief summary of our methodology 
below. For a much more complete description see Kauffmann et al. (2003a).

The rich stellar absorption-line spectrum of a typical SDSS galaxy is both a 
blessing and a curse. While the lines provide unique information about the 
stellar content and galaxy dynamics, they make the measurement of weak nebular 
emission lines quite difficult. To deal with this, we have performed a careful 
subtraction of the stellar absorption-line spectrum before measuring the 
nebular emission lines. This is accomplished by fitting the emission-line-free 
regions of the spectrum with a model galaxy spectrum. The model spectra are 
based on the new population synthesis code of Bruzual \& Charlot (2003), which 
incorporates high-resolution stellar libraries. A set of 39 model template 
spectra were used that span the relevant range in age and metallicity.  After 
convolving the template spectra to the measured stellar velocity dispersion 
($\sigma_*$) of an individual SDSS galaxy, the best fit to the galaxy spectrum 
is constructed from a linear combination of the template spectra.

As diagnostics of the stellar population we have used the amplitude of the 
4000 \AA\ break [the $D_n(4000)$ index of Balogh et al. 1999] and the strength 
of the H$\delta$ absorption line (the Lick H$\delta_A$ index of Worthey \& 
Ottaviani 1997). The diagnostic power of these indices is shown in Figure 1.1 
above.  In both cases the indices measured in the SDSS galaxy spectra are 
corrected for the flux of the emission lines in their bandpasses.

Using a library of 32,000 star formation histories that span the relevant 
range in metallicity, we have used the measured $D_n(4000)$ and H$\delta_A$ 
indices to estimate the SDSS $z$-band mass-to-light ratio for each galaxy.
By comparing the colors of our best-fitting model to those of each galaxy,
we have estimated the dust extinction of starlight in the $z$-band
(e.g., Calzetti, Kinney, \& Storchi-Bergmann 1994; Charlot \& Fall 2000).

The SDSS imaging data are then used to provide the basic structural parameters.
The $z$-band absolute magnitude and the derived values of $M/L$ and $A_z$
yield the stellar mass ($M_*$). The half-light radius in the $z$-band and the
stellar mass yield the effective stellar surface mass density
($\mu_* = M_*/2\pi r_{50,z}^2$). As a proxy for Hubble type we use
the SDSS ``concentration'' parameter $C$, which is defined as the ratio
of the radii enclosing 90\% and 50\% of the galaxy light in the $r$ band
(see Stoughton et al. 2002). Strateva et al. (2001) find that galaxies
with $C >$ 2.6 are mostly early-type galaxies, whereas spirals and irregulars
have 2.0 $< C <$ 2.6.

\subsection{Results}

As discussed in  Kauffmann et al. (2003b - hereafter K03), the overall SDSS 
galaxy population is remarkably bimodal in nature.  There is a rather abrupt 
transition in properties at a critical stellar mass 
$M_* \approx 3 \times 10^{10} M_{\odot}$. Below this mass, galaxies are young
[small $D_n(4000)$ and large H$\delta_A$], are disk dominated (low 
concentration, $C <$ 2.6), and show a strong increase in surface mass density 
($\mu_*$) with increasing mass. Above this mass, galaxies are old [large 
$D_n(4000)$ and small H$\delta_A$], are bulge dominated ($C >$ 2.6), and have 
a uniform $\mu_*$. There is also a strong transition in mean stellar age at 
a characteristic surface mass density of $3 \times 10^8 M_{\odot}$ kpc$^{-2}$ 
and a concentration index of 2.6. 

Where do AGN hosts fit into this landscape? We find that AGNs are commonly 
present only in galaxies with large masses ($>10^{10} M_{\odot}$), stellar 
velocity dispersions ($>100$ ~km~s$^{-1}$), and surface mass densities  
($>10^8 M_{\odot} {\rm kpc}^{-2}$)\footnote{Tremonti et al. (in preparation) 
find the SDSS galaxies obey a strong mass-metallicity relation. Since AGNs are 
hosted by massive galaxies, they have high metallicity. This explains why the 
``AGN plume'' ascends from the region of metal-rich starformers in Figure 
1.2.}.  Their stellar content and structure varies significantly as a 
function of AGN luminosity.

Before proceeding, a brief digression is necessary.  Any attempt to 
characterize the stellar population in the AGN hosts with these spectra must, 
by necessity, include the transition objects that comprise the majority of the 
AGNs in Figure 1.2. Excluding these would bias the sample against host galaxies
with significant amounts of on-going star formation. The only disadvantage
of including the transition class is that it is then difficult (on the basis of
the emission-line ratios alone) to classify them as Seyferts with star 
formation or LINERs with star formation.

It has long been known that LINERs do not attain the high luminosities of 
powerful Seyfert nuclei (e.g., Heckman 1980b; Ho et al. 2003)\footnote {In the 
SDSS sample, the luminosity of the [O~III] $\lambda$5007 emission line (the 
strong line from the NLR that is least contaminated by a contribution from 
galactic H~II regions) is larger in type 2 Seyferts than in LINERs by an 
average factor of about 30 to 100 for galaxies with the same stellar mass or 
velocity dispersion.  If we assume that AGNs hosts obey the (in)famous 
relation between black hole mass and stellar velocity dispersion, this would 
imply that on average LINERs are AGNs operating in a mode with a much lower 
$L/L_{\rm Edd}$ than Seyfert nuclei.  The same trend is seen by Ho (2002, 
2003) in nearby galaxies}.  Thus, rather than classifying AGNs according to 
their line ratios, it makes more sense to examine how the properties of AGN 
hosts vary as a function of AGN luminosity (using the [O~III] $\lambda$5007 
line as a measure). This allows us to include the transition class AGNs and 
thus mitigate the selection effects described above. For reference, the 
change-over from a LINER-dominated to Seyfert-dominated population occurs at 
an extinction-corrected value of $L_{\rm [O~III]} \approx 10^7 \,L_{\odot}$
in the SDSS sample.

\begin{figure}
\centering
\includegraphics[height=140mm,width=120mm]{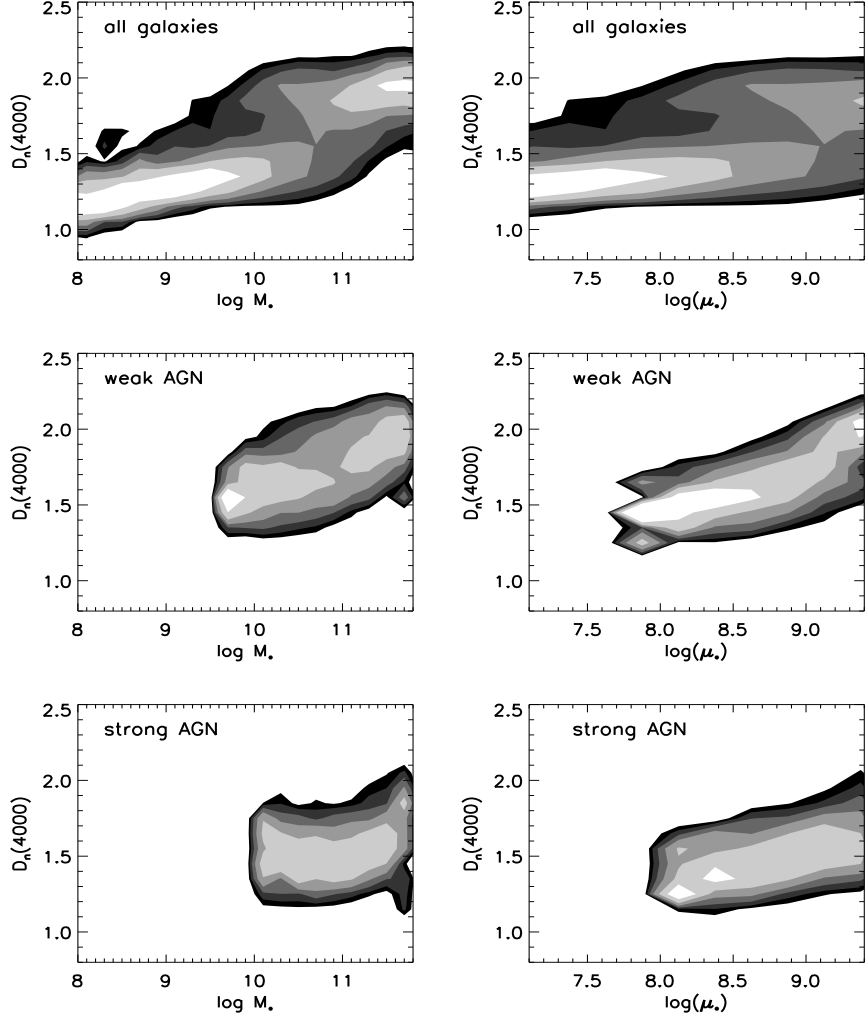}
\caption{Contour plots of conditional density distributions (see K03) showing 
trends in the age-sensitive index $D_n(4000)$ as a function of stellar mass 
$M_*$ and surface mass density $\mu_*$ for all galaxies ({\it top}), weak AGNs 
with $L_{\rm [O~III]} < 10^7 \,L_{\odot}$ ({\it middle}), and powerful AGNs 
with $L_{\rm [O~III]} > 10^7 \,L_{\odot}$ ({\it bottom}). Galaxies have been 
weighted by $1/V_{max}$, and the bivariate distribution function has been 
normalized to a fixed number of galaxies in each bin of log $M_*$ ({\it left}) 
and log $\mu_*$ ({\it right}). The plots are only made over ranges in $M_*$ 
and $\mu_*$ where there are at least 100 objects per bin (so the plots for the 
AGN hosts do not extend to low $M_*$ or $\mu_*$).}
\end{figure}

We find that the stellar content of the AGN hosts is a strong function of 
$L_{\rm [O~III]}$ (Fig. 1.3).  This figure shows that the galaxy population 
as a whole is strongly bimodal (young, low mass, and low density, {\it or} 
old, high mass, and high density).  What is very striking is that while the 
weak AGNs roughly follow the same trend as the normal massive galaxies, the 
powerful AGNs do not. The powerful AGNs are hosted by galaxies that are 
massive and dense, but relatively young. 

Could the relatively small value for the $D_n(4000)$ index be caused by 
featureless AGN continuum rather than young stars? Decisive evidence against 
this is provided by the Balmer absorption lines, which are strong in the hosts 
of powerful AGNs. This is illustrated in Figure 1.4 which compares a high 
signal-to-noise ratio composite spectrum of several hundred of the most 
powerful AGNs ($L_{\rm [O~III]} > 10^8 \,L_{\odot}$) to a similar composite 
spectrum of normal high-mass, star-forming galaxies. More generally, nearly 
all the powerful AGNs in our sample lie on or above the locus of the 
continuous models of star formation in Figure 1.1.  They have strong Balmer 
absorption lines as well as a relatively small $D_n(4000)$ index, so the 
latter is not due to dilution by AGN light.

\begin{figure}
\centering
\includegraphics[height=90mm,width=110mm]{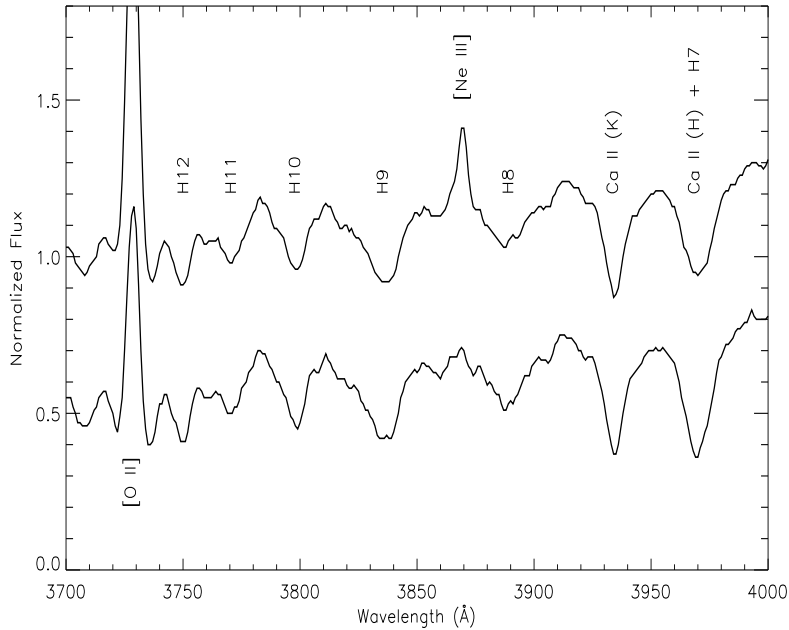}
\caption{Comparison of composite near-UV spectra for the hosts of several 
hundred of the most powerful type 2 Seyferts in the SDSS ({\it top}) with a 
similar composite spectrum of normal star-forming galaxies with similar 
stellar masses as the Seyferts ({\it bottom}). Both spectra have been 
normalized to unit flux at 4200 \AA\ and then the composite spectrum for the 
normal galaxies has been offset by $-$0.5 flux units for clarity. Note
that both spectra have similar strengths for the Ca~II lines from an old 
stellar population and the high-order Balmer absorption lines from young stars.}
\end{figure}

K03 found that the fraction of galaxies that have undergone a major burst of 
star formation within the past Gyr declines strongly with increasing galaxy 
mass. The fraction of all SDSS galaxies that have experienced a major burst 
(at better than the 97.5\% confidence level) is only about 0.1\% to 0.3\% over 
the range in $M_*$ appropriate for typical AGN hosts. In contrast, the fraction
of ``high-confidence'' bursts in the AGN hosts rises with increasing AGN 
luminosity, from $\sim$1\% at the lowest luminosities to nearly 10\% at the 
highest.

Thus, it appears that powerful type 2 AGNs (Seyferts) reside in galaxies that 
have a combination of properties that are rare in the galaxy population as 
a whole: they are massive (and dense) but have relatively young stellar 
populations. This may have a very simple explanation. The two necessary 
ingredients for a powerful AGN are a massive black hole and an abundant fuel 
supply. Only massive galaxies contain massive black holes, and only galaxies 
with significant amounts of recent/ongoing star formation have the requisite 
fuel supply (ISM). This combination is rare today. The most massive black 
holes live mostly in the barren environment of a massive early-type galaxy. 
This was evidently not the case during the AGN era.

The rarity of galaxies having this combination of properties today implies 
that we may be witnessing a transient event.  This would be consistent with
the relatively large fraction of the hosts of powerful AGNs that have 
undergone a major starburst within the last $\sim 1$ Gyr (a rough galaxy 
merger time scale).

The above picture is appealingly simple. However, one possible ``fly in the 
ointment'' is the recent result that the hosts of powerful low-redshift QSOs 
are very massive but otherwise normal elliptical galaxies (Dunlop et al. 
2003). If the stellar population of the QSO hosts is in fact old, then our 
results on the hosts of powerful Seyfert 2s could mean that the standard
unified scenario for AGNs is seriously incomplete at high luminosities.

\section{Summary}

I have reviewed our understanding of the connection between star formation and 
the AGN phenomenon, emphasizing large surveys that have used direct 
spectroscopic diagnostics to search for young ($< 1$ Gyr) stars in AGNs and in 
their host galaxies in the local Universe.  I have considered only those 
objects in which the blinding glare of the central AGN is obscured from our 
direct view, so that the optical continuum is dominated by starlight. For 
these ``type 2'' AGNs I have used the luminosity of the narrow-line region as 
an indicator of AGN power. 

The major conclusions are as follows:

{\it Galactic Nuclei ($\sim10^2$ to $10^3$ pc):}
The majority of galaxies of early/middle Hubble type contain AGNs, usually of 
low luminosity. In most cases, the nuclear stellar population in low-power 
AGNs is predominantly old. In contrast, a young stellar population is detected 
in about half of the powerful type 2 Seyfert nuclei.

{\it Host Galaxies ($\sim10^4$ pc):}
The host galaxies of AGNs have large stellar masses and surface mass 
densities. Normal galaxies like this have predominantly old stellar 
populations. While the stellar content of the hosts of low-power AGNs appears
normal, the hosts of powerful AGNs have young stellar populations. A 
significant fraction of these young hosts have undergone major starbursts in 
the last $\sim 1$ Gyr.

The observational evidence for a connection between star formation and black 
hole fueling is thus clearer in powerful AGNs. This is true both in the 
nucleus itself and in its host galaxy. Such a dependence is astrophysically 
plausible, and is qualitatively consistent with the processes that established 
the correlation between black hole and stellar mass in galaxy bulges.

The most natural explanation for why powerful AGNs are hosted by massive 
galaxies with young stellar populations is that a powerful AGN requires both 
a high mass black hole (massive host galaxy) and an abundant fuel supply (a 
star-forming ISM). While this combination is rare today, it would have been 
much more common during the AGN era at high redshift.\\[3mm]

I would like to thank Carnegie Observatories, in particular Luis Ho, for 
organizing and hosting such a successful meeting. I would also like to thank 
all my collaborators on the work reported in this paper, with a special thanks 
to Guinevere Kauffmann who is leading the SDSS project on AGN hosts.

Funding for the creation and distribution of the SDSS Archive has been provided
by the Alfred P. Sloan Foundation, the Participating Institutions, the National
Aeronautics and Space Administration, the National Science Foundation, the U.S.
Department of Energy, the Japanese Monbukagakusho, and the Max Planck Society.
The SDSS Web site is http://www.sdss.org/.

The SDSS is managed by the Astrophysical Research Consortium (ARC) for the
Participating Institutions. The Participating Institutions are The University
of Chicago, Fermilab, the Institute for Advanced Study, the Japan Participation
Group, The Johns Hopkins University, Los Alamos National Laboratory, the
Max-Planck-Institute for Astronomy (MPIA), the Max-Planck-Institute for
Astrophysics (MPA), New Mexico State University, University of Pittsburgh,
Princeton University, the United States Naval Observatory, and the University
of Washington.

\end{document}